\documentclass[a4paper,fleqn]{cas-dc}
\usepackage[numbers]{natbib}
\usepackage{graphicx}  
\usepackage{enumitem}
\usepackage{xurl}
\usepackage{listings}
\usepackage{xcolor}
\usepackage{float}

\usepackage{subcaption}
\lstdefinelanguage{Solidity}{
  morekeywords={
    contract, event, address, uint256, external, function
  },
  sensitive=true,
  morecomment=[l]{//},
  morecomment=[s]{/*}{*/},
  morestring=[b]",
}

\def\tsc#1{\csdef{#1}{\textsc{\lowercase{#1}}\xspace}}
\tsc{WGM}
\tsc{QE}
\tsc{EP}
\tsc{PMS}
\tsc{BEC}
\tsc{DE}

\begin{document}
\let\WriteBookmarks\relax
\def\floatpagepagefraction{1}
\def\textpagefraction{.001}

\shorttitle{A Secure Standard for NFT Fractionalization}
\shortauthors{Haouari et~al.}

\title [mode = title]{A Secure Standard for NFT Fractionalization}                      

\newlist{noitemsepitemize}{itemize}{1}
\setlist[noitemsepitemize,1]{label=-, nosep, left=0pt, topsep=0pt, partopsep=0pt, itemsep=0pt, parsep=0pt}


%
\author[1]{Wejdene Haouari}[type=editor,orcid=0000-0003-2960-8629]
\cormark[1]
\cortext[cor1]{Corresponding author}
\ead{wejdene.haouari1@gmail.com}


\credit{Conceptualization of this study, Methodology, Software}

\affiliation[1]{organization={Department of Electrical Engineering \& Computer Science, York University, ON, Canada},
    }

\author[1]{Marios Fokaefs}[style=chinese]



\begin{abstract}
Non-fungible tokens (NFTs) offer a unique method for representing digital and physical assets on the blockchain. However, the NFT market has recently experienced a downturn in interest, mainly due to challenges related to high entry barriers and limited market liquidity. Fractionalization emerges as a promising solution, allowing multiple parties to hold a stake in a single NFT. By breaking down ownership into fractional shares, this approach lowers the entry barrier for investors, enhances market liquidity, and democratizes access to valuable digital assets. Despite these benefits, the current landscape of NFT fractionalization is fragmented, with no standardized framework to guide the secure and interoperable implementation of fractionalization mechanisms.
This paper contributions are twofold: first, we provide a detailed analysis of the current NFT fractionalization landscape focusing on security challenges; second, we introduce a standardized approach that addresses these challenges, paving the way for more secure, interoperable, and accessible NFT fractionalization platforms.

\end{abstract}




\maketitle

\section{Introduction}

In the rapidly evolving landscape of blockchain technology, Non-Fungible Tokens (NFTs) have emerged as a significant innovation, transforming our understanding of digital ownership and asset management \cite{wang2021nonfungible}. NFTs, unique digital assets verified using blockchain, have captured the attention of various industries, ranging from digital art to real estate. However, despite their initial popularity, the NFT market has recently encountered challenges, including a noticeable decline in interest and questions about their long-term viability and accessibility. This downturn can be attributed to several factors, such as market saturation and the high cost of acquiring top-tier NFTs. These challenges highlight the need for innovative solutions to revitalize the market and broaden its appeal.

One of the primary problems that led to the decline of the NFT market, is the limited accessibility due to the high cost associated with acquiring unique and valuable NFTs. The exclusivity of ownership has created barriers, preventing a broader audience from participating in the market. 3

NFT fractionalization has emerged as a promising solution to address these issues. By enabling shared ownership of digital assets, fractionalization lowers the entry barrier for potential investors  \cite{Wejdene}. This concept democratizes digital asset ownership, making NFTs more accessible to a broader audience and potentially revitalizing the market by introducing NFTs to new sectors. However, despite its potential, the current landscape of NFT fractionalization lacks uniformity, resulting in various implementations with differing security and functional qualities. This lack of standardization raises barriers to entry due to learning curves, limits interoperability between platforms, and poses significant security risks.

This paper aims to fill the gap by analyzing existing NFT fractionalization solutions, identifying their limitations, and proposing a standardized framework to enhance security and interoperability. Our contributions include a detailed security analysis of current solutions, the proposal of a standardization framework for fractionalization, and the development of a concrete implementation that adheres to this framework.

The structure of this paper is as follows: Section \ref{Background} provides a background on the relevant standards. Section \ref{Solution} reviews current NFT fractionalization platforms, analyzing their strengths and weaknesses.  Section \ref{proposoal} introduces our standardization proposal. In Section \ref{concrete}, we detail the development of a concrete implementation based on our proposed standard. Section \ref{Related} examines related work in the field. Section \ref{future}  offers a discussion of the findings and directions for future work. Finally, Section \ref{Conclusion} concludes the paper by summarizing the contributions.
\section{Background}
\label{Background}

\subsection{Standardization in Blockchain Technology}

In the Ethereum ecosystem, smart contract standards are crucial in defining how contracts should operate and interact. To bring consistency and interoperability among various contracts, the Ethereum community has established a set of standards known as Ethereum Requests for Comments (ERCs). These are part of the broader category of Ethereum Improvement Proposals (EIPs) \cite{eip}.

\textbf{ERC-20: Token Standard}
\\ERC-20 is a standard for creating and managing tokens on the Ethereum blockchain \cite{erc20}. Established by Fabian Vogelsteller, it has been instrumental in the growth of Initial Coin Offerings (ICOs) and decentralized finance (DeFi) applications. ERC-20 tokens are fungible, meaning each token is identical and interchangeable with others, similar to conventional currencies or assets like gold. This standard ensures that tokens adhering to the ERC-20 interface are mutually compatible, facilitating their integration across various wallets, exchanges, and decentralized applications.

\textbf{ERC-721: Non-Fungible Token Standard}
\\The ERC-721 standard, a cornerstone for non-fungible tokens (NFTs) on Ethereum, provides a framework for the creation, management, and trading of unique digital assets \cite{erc721}. This standard defines a set of functions and events that enable consistent interaction with NFTs. Functions like \texttt{safeTransferFrom} allow for secure transfers of NFTs between addresses, while \texttt{approve} grants third-party transfer permissions. Meanwhile, the \texttt{balanceOf} function reports the quantity of NFTs held by a particular address. ERC-721's unique feature is the ability to represent individual, non-interchangeable assets, making it ideal for digital collectibles, art, and more.

\textbf{ERC-1155: Multi Token Standard}
\\Developed by Enjin Radomski et al. \cite{erc1155}, ERC-1155 is a versatile smart contract standard capable of managing multiple types of tokens within a single contract. This standard supports fungible (like ERC-20) and non-fungible (like ERC-721) tokens, offering a unified approach to token management. This innovation simplifies the process and enhances efficiency in various applications, including gaming, digital collectibles, and artwork.

\textbf{ERC-165: Interface Detection Standard}
\\ ERC-165 specifies a standard method for Ethereum contracts to announce and identify the interfaces they support  \cite{erc165}. This standard is essential for enhancing interoperability among contracts and applications. It allows contracts to query whether a given contract implements a specific interface using the \texttt{supportsInterface} function. This capability is handy for developers creating modular and flexible applications, as it enables them to easily determine the functionalities a contract supports and interact with it accordingly.

While the existing ERC standards have provided a solid foundation for various blockchain applications, they present significant limitations regarding NFT fractionalization. Currently, there is no standard specifically designed for NFT fractionalization. Existing standards like ERC-721 and ERC-1155 are not equipped to handle the complexities of fractional ownership, leaving developers to create ad-hoc solutions that may lack consistency and security. Given this, there is a clear need for a new standard specifically designed to address the challenges of fractionalizing NFTs.

\subsection{Fractionalization}

NFT fractionalization, often referred to as Fractional NFTs (F-NFTs) \cite{Wejdene}, is the process of breaking down a single Non-Fungible Token (NFT) into smaller, manageable parts known as \emph{fractions} or \emph{shares}. These fractions allow for the shared ownership of an NFT among multiple individuals, thus democratizing access to otherwise prohibitively expensive assets. Each fraction represents a tokenized stake in the NFT, enabling individual investors or collectors to buy and sell their shares independently.
This approach is especially beneficial for high-value NFT assets such as exclusive art, luxury items, or real estate, which typically would be unaffordable for an average individual. Through fractionalization, these assets become financially accessible as investors can purchase just a portion of the NFT at a lower cost.
Moreover, the fractionalization of NFTs extends the scope of these digital assets beyond traditional domains, such as digital art, to more tangible sectors like real estate and luxury goods. It also opens up new investment opportunities and scenarios, including in emerging fields like decentralized finance, thus broadening the potential and reach of NFTs in various markets.

\subsection{Decentralized Finance (DeFi)}

Decentralized Finance (DeFi) simplifies finance by using technology to manage financial activities like lending, borrowing, and trading without central authorities. It relies on smart contracts on blockchains, mainly Ethereum, to operate transparently and without intermediaries \cite{Mohan2022}.
Liquidity pools, essential components of DeFi, serve as token reserves within smart contracts, enabling direct exchanges of assets or loans. These pools allow users to earn fees proportionate to their contribution. An example of this mechanism in action is Uniswap \cite{sushiswap}, where users can swap various tokens based on the supply available in the pool, eliminating the necessity for a direct match between buyers and sellers. Furthermore, DeFi leverages the concept of staking \cite{Mohan2022}, where users lock up their tokens to support network operations in return for rewards, often contributing to liquidity pools or participating in governance decisions. Additionally, yield farming emerges as a strategy for maximizing returns by reallocating assets across different protocols and engaging in lending or staking activities to earn interest or rewards.

NFT fractional tokens can be integrated into DeFi platforms, which are used similarly to other digital assets. Owners of fractional tokens can engage in liquidity provision, staking, and collateralization, opening new avenues for investment and earning within the DeFi space. 
\section{Current Solutions}
\label{Solution}
\subsection{Tessera}





\begin{figure*}
    \centering
     \includegraphics[width=0.8\linewidth]{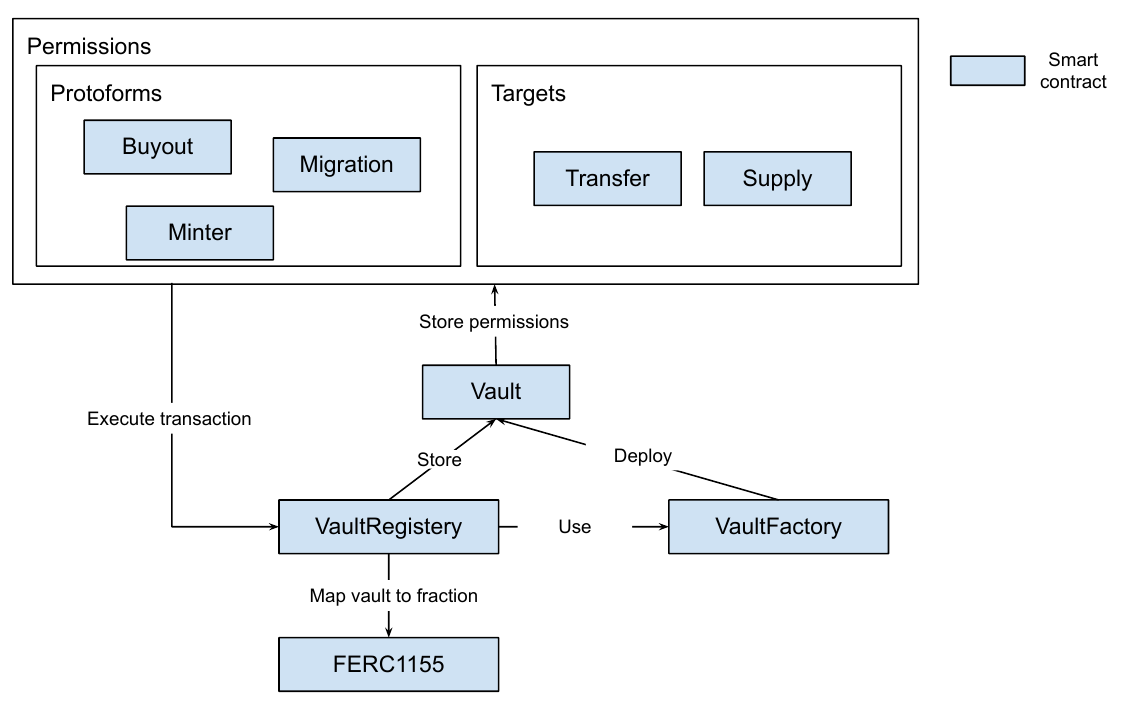}

   \caption{Tessera platform smart contracts }
       \label{fig:VaultInteractionDiagram}

\end{figure*}

Tessera, launched in August 2022, introduces a novel approach to owning and governing Non-Fungible Tokens exclusively on the Ethereum blockchain  \cite{Tessera-medium}. The platform’s architecture is based on the concept of Hyperstructures \cite{Hyperstructures}, which are blockchain protocols designed to operate autonomously, eliminating the need for continuous maintenance, intermediaries, or centralized control. Tessera's structure revolves around several key components, our analysis will concentrate on nine primary contracts \cite{tesseraGithub}. Figure \ref{fig:VaultInteractionDiagram} shows the interaction between these smart contracts.

\begin{enumerate}
  \item \textbf{Vault Creation: }  The \textit{Vault} Contract is central to Tessera’s functionality, employing a Merkle Tree-based permission system \cite{tesseraGithub} to ensure security while allowing for post-deployment modifications. Vault creation is facilitated by the \textit{VaultFactory} contract, leveraging cloning technology for efficiency. The \textit{VaultRegistry} maintains a record of vaults and their fractional ownership, represented by FERC1155 tokens, ensuring transparency and security.
  \item \textbf{Modules and Targets:} Permissions in Tessera are managed through a combination of Modules and Targets. Modules are advanced contracts that offer functionalities such as auction-based ownership transfers, token issuance with a fixed supply, and transitions of vaults to new governance models. Targets, on the other hand, are simpler contracts executed by the vault to manage the lifecycle of tokens, including minting, burning, and transferring. This modular approach allows Tessera to offer flexibility and programmability in managing fractionalized NFTs.
\end{enumerate}

     Despite its innovative architecture, Tessera struggles with integration issues with broader DeFi platforms and utilizes ERC1155 tokens which, while versatile, limit broader market adoption compared to ERC20 tokens.
     Security vulnerabilities and the complexity of managing a hyperstructured system pose ongoing challenges, impacting user trust and system reliability. Ultimately, these issues may have contributed to the Tessera platform announcing its shutdown on September 1st, 2023 \cite{Tesseradown}. The decision was driven by financial challenges and an unsustainable economic model.

\subsection{Unic.ly}

\begin{figure*}
    \centering
     \includegraphics[width=0.8\linewidth]{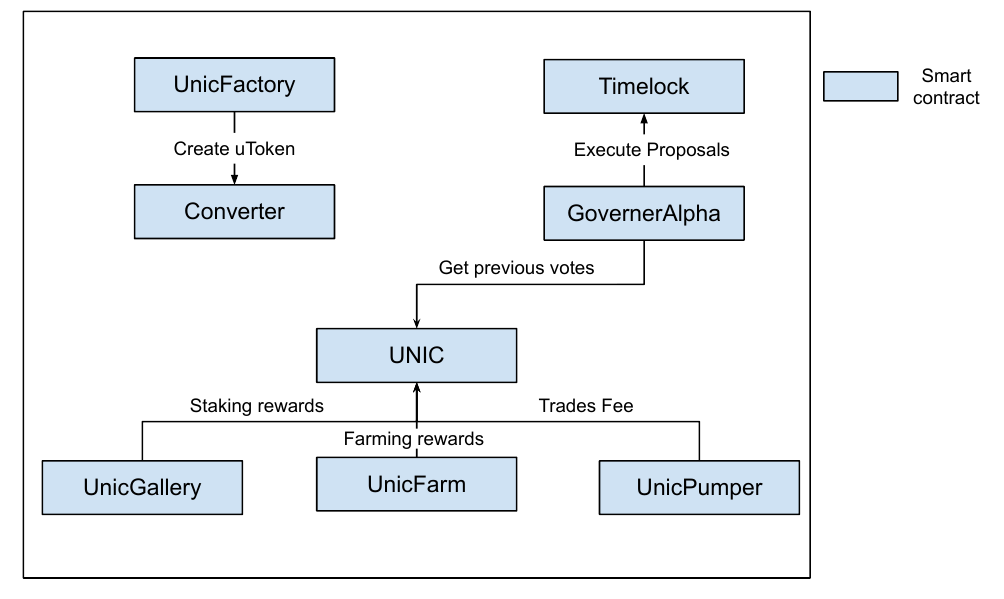}

   \caption{Unic.ly platform smart contracts }\label{fig:unic-interaction}
\end{figure*}

Unic.ly, launched in November 2021, is a decentralized  NFT system that allows users to tokenize their assets and engage in trading while including DeFi principles such as staking, yield farming, and liquidity mining \cite{unicly}. Unic.ly distinguishes itself through user-friendly design and architecture, making it accessible to a broad audience. Unic.ly is a platform for manufacturing digital tokens and offering optimal prices for traders, with an NFT creator and an automated market maker (AMM) enabled marketplace \cite{mohan2022automated}.
Figure \ref{fig:unic-interaction} illustrates the interaction among the primary smart contracts.

\begin{enumerate}
  \item \textbf{uToken Creation and NFT Fractionalization: } \textit{UnicFactory} initiates the process by minting uTokens, which represent fractional ownership of NFTs. The \textit{Converter} contract fractionalizes these NFTs into uTokens and facilitates auctions for these fractionalized assets. Users can buy, sell, or trade fractions of NFTs, and to redeem an underlying NFT from the collection, uToken holders must first vote to unlock the collection.
  \item \textbf{Staking and Earning Rewards:} \textit{UnicGallery} offer a platform for users to stake their uTokens and earn UNIC governance tokens as rewards, incentivizing users to participate in the ecosystem. Similarly, \textit{UnicFarm} allows users to stake liquidity provider (LP) tokens from various pools to earn UNIC rewards.
  \item \textbf{Governance and Platform Decisions: }UNIC tokens play a pivotal role in the platform’s governance. Holders of UNIC tokens can participate in governance decisions by proposing changes or voting on proposals through the \textit{GovernorAlpha} contract, with the \textit{Timelock} contract adding a mandatory delay to the execution of governance actions.
  \item \textbf{Fee Conversion and Reward Distribution:} The \textit{UnicPumper} contract converts tokens collected as fees into UNIC tokens, which are then distributed back into the ecosystem as rewards for staking.
\end{enumerate}
 
\subsection{NFTX}

Introduced in December 2020, NFTX is a platform designed to create fluid markets for otherwise illiquid Non-Fungible Tokens \cite{nftxDoc}. Users can deposit their NFTs into an NFTX vault, and in return, they generate a fungible ERC20 token, known as a vToken. This vToken signifies a claim on an arbitrary asset within the vault. Furthermore, vTokens can be used to reclaim a particular NFT from a vault.

 \begin{figure*}
    \centering
     \includegraphics[width=0.9\linewidth]{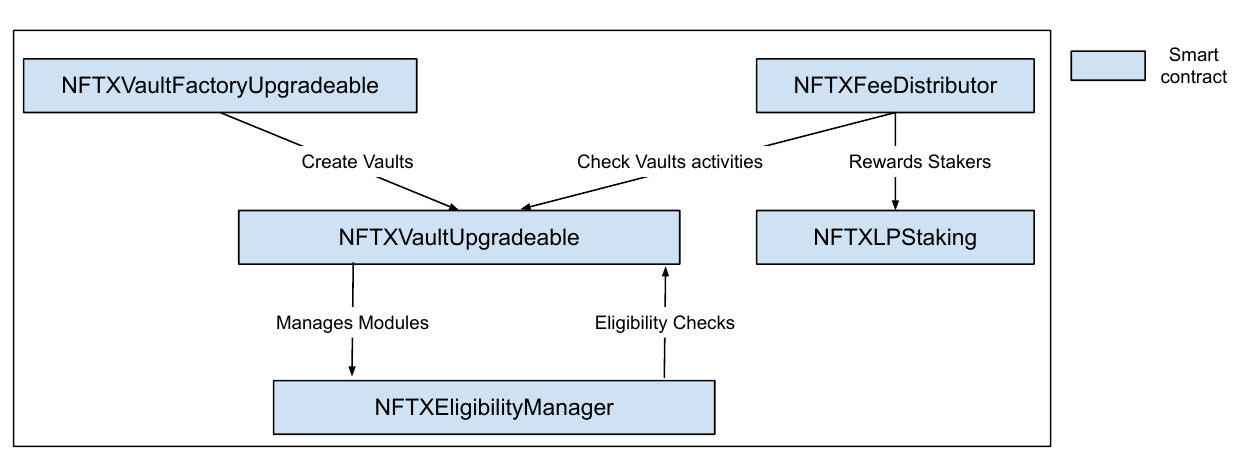}

   \caption{NFTX platform smart contract }\label{fig:nftx-interaction}
\end{figure*}

Figure \ref{fig:nftx-interaction} illustrates the interaction among the primary smart contracts.

\begin{enumerate}
  \item \textbf{Vault Creation and Configuration:} Users can create vaults for any NFT collection by specifying the NFT asset address, which points to the smart contract of the NFTs. This process creates a new ERC-20 token (vToken) for that vault. Vault creators can set eligibility criteria to specify which NFTs can be minted into the vault and configure features like minting, random redeems, and targeted redeems.
  \item \textbf{Minting Process:} NFT owners can mint vTokens by depositing their NFTs into a corresponding vault. Each NFT minted into the vault increases the supply of vTokens by one, minus any mint fee set by the vault creator.
  \item \textbf{Redeeming NFTs:} vToken holders can redeem a random NFT from the vault by spending the corresponding vTokens value. For an additional fee, users can select a specific NFT they wish to redeem from the vault.
  \item \textbf{Staking and Rewards:} Users can provide liquidity by adding equal values of vTokens and a base currency (e.g., ETH) to a liquidity pool on an Automated Market Maker (AMM) like SushiSwap. In return, they receive LP tokens representing their share of the pool. LP tokens can be staked in the NFTX platform to earn a portion of the fees generated by vault activities.
\end{enumerate}

\subsection{Comparative Analysis of Current Solutions}
\textbf{Shared Functions and Architecture:}

Tessera, Unic.ly, and NFTX share core functionalities, particularly in vault creation, tokenization, and governance, albeit with distinct implementations:

\begin{itemize}
  \item \textbf{Vault Mechanism:} All platforms use a vault smart contract to store NFTs and issue fractional tokens. 
  \item \textbf{Tokenization:} Fractional ownership is represented through tokens issued upon NFT deposit. Tessera uses the FERC1155 standard, while Unic.ly and NFTX utilize ERC20 tokens. Despite differing token standards, all platforms enable trading of fractional ownership.
  \item \textbf{Governance:} Each platform includes governance mechanisms that allow token holders to influence decisions. The complexity and depth of these mechanisms differ, with Tessera offering a more complex governance model than the others.
  \item \textbf{Auction Systems:} All platforms facilitate ownership transfer through auctions. Tessera embeds its auction system within its governance structure, Unic.ly simplifies the process for users, and NFTX emphasizes liquidity pools and token redemption for exchanges.
\end{itemize}

\textbf{Unique Features:}

Although the platforms share a common framework, several unique features distinguish their approaches:

\begin{itemize}
  \item \textbf{Tessera’s Modular Architecture:} Tessera employs a highly modular system with multiple smart contracts handling different functions such as auction transfers, fractional token minting, and governance transitions. While this flexibility suits advanced use cases, it introduces complexity and potential security risks. Tessera’s use of the FERC1155 token limits interoperability with ERC20-based DeFi platforms, and its lack of direct DeFi integration reduces its appeal to liquidity-focused users.
  \item \textbf{Unic.ly’s Simplicity and DeFi Integration:} Unic.ly prioritizes simplicity, using ERC20 tokens for broad compatibility with wallets and decentralized applications. Its strong integration with DeFi features, including staking, yield farming, and liquidity mining, makes it attractive to users seeking to participate in both NFT fractionalization and DeFi activities. However, this simplicity limits the flexibility in managing fractionalized assets compared to Tessera.
  \item \textbf{NFTX’s Liquidity and Market Making:} NFTX excels at creating liquidity for NFTs.  While NFTX’s design supports decentralized exchanges and liquidity provision, it lacks the advanced governance features and asset management flexibility offered by Tessera.
\end{itemize}

\section{NFT fractionalization standard} 
\label{proposoal}
This section defines a standardization proposal for the fractionalization of NFTs.
The proposal revolves around defining a set of smart contracts, each with specific roles and interactions, to ensure a secure, standardized, and interoperable framework across different platforms and applications that serve as a baseline for future fractionalization platforms. This standard proposal is an improved version of the one presented in our previous paper \cite{Wejdene}, enhancing security and efficiency while incorporating a governance mechanism, a more modular approach, and DeFi integration to provide a reliable foundation for future fractionalization solutions.

\begin{figure*}
    \centering
    \includegraphics[width=0.9\linewidth]{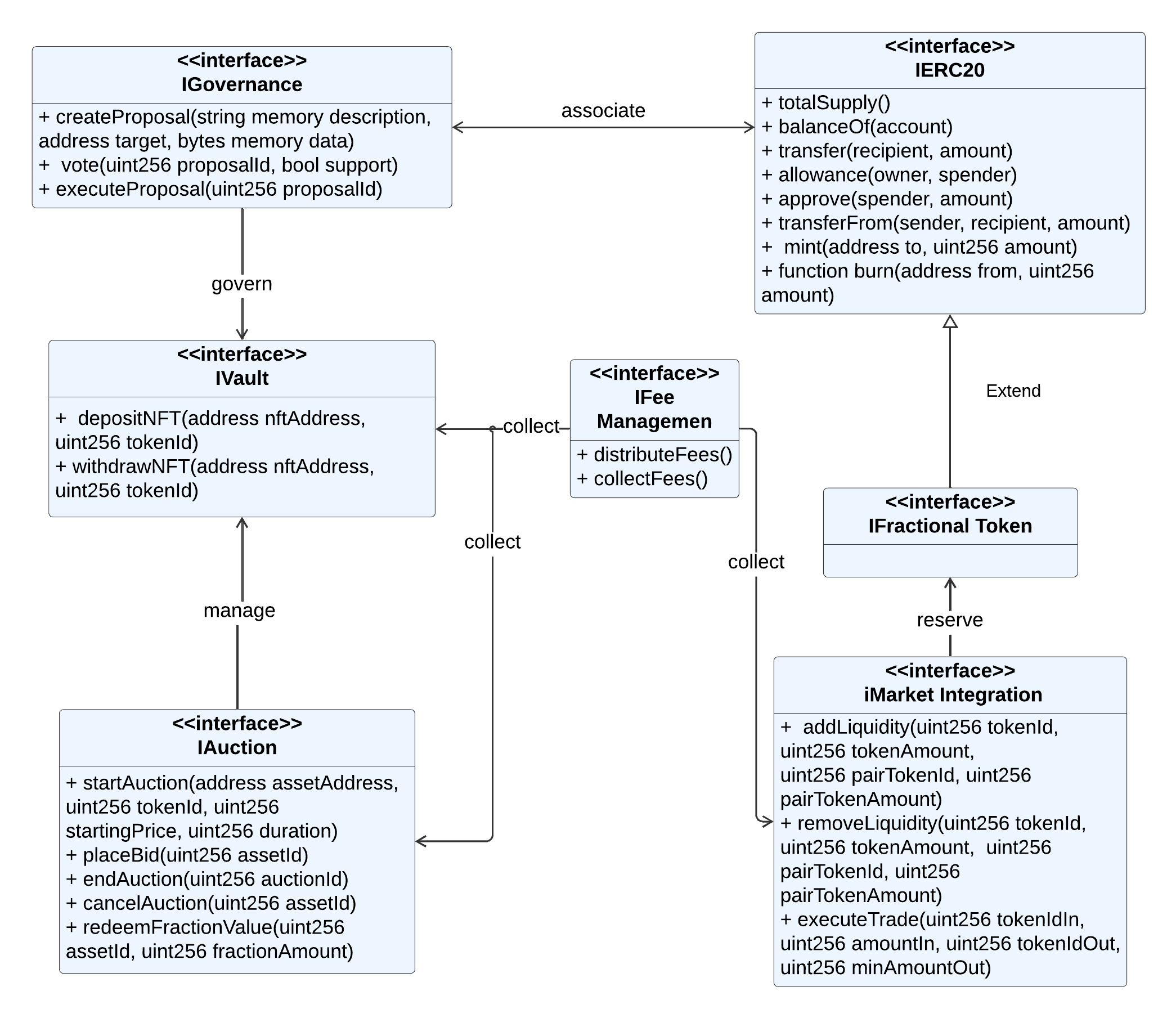}
    \caption{Standardization proposal Interfaces}
    \label{fig:Interfaces}
\end{figure*}

Figure \ref{fig:Interfaces} displays the class diagram for our proposed solution, which is further elaborated in the subsequent sections.

After analyzing the existing solutions, it becomes evident that most of them share similar roles in the fractionalization process. This typically involves several key smart contracts, including the Vault Contract, Fractional Token Contract, Auction Contract, Governance Contract, Automated Market Maker (AMM) Contract, and Fee Management Contract.

\subsection{ Vault Contract Interface}

The Vault Contract Interface defines the essential functions for securely managing the lifecycle of NFTs within the fractionalization process. It outlines standardized methods for depositing NFTs, minting fractional tokens, and withdrawing NFTs, ensuring interoperability and consistent behavior across different implementations.

\textbf{a. NFT Deposit Function:}

This function transfers an NFT from the depositor's address to the vault. It must verify the caller's ownership of the NFT and initiate the minting of fractional tokens corresponding to the NFT's value.

\begin{itemize}
    \item \textbf{nftAddress:} This parameter specifies the address of the NFT contract. Identifying the specific ERC-721 or ERC-1155  contract that manages the NFT intended for deposit. By providing this address, the Vault Contract can interact directly with the NFT contract to transfer ownership of the NFT into the vault.
    \item \textbf{tokenId:} The unique identifier for the NFT within its contract. The tokenId, with the nftAddress, uniquely identifies the NFT across the Ethereum blockchain.
\end{itemize}

\begin{lstlisting}[language=Solidity]

function depositNFT(address nftAddress, uint256 tokenId) external;
\end{lstlisting}

\textbf{b. NFT Withdrawal Function:}

This function facilitates the withdrawal of an NFT from the vault, enabling fractional token holders to claim the original NFT. The contract must define specific criteria for withdrawal, such as owning a majority of the fractional tokens or auction mechanism.

\begin{itemize}
    \item \textbf{nftAddress:} It is essential to specify the NFT contract from which an NFT will be withdrawn, just like in the deposit function. This ensures that the correct NFT contract is targeted during the withdrawal operation.
    \item \textbf{tokenId:} This parameter specifies the NFT to be withdrawn based on its unique identifier. It ensures correct transfer based on fractional token ownership.
\end{itemize}

\begin{lstlisting}[language=Solidity]

function withdrawNFT(address nftAddress, uint256 tokenId) external;
\end{lstlisting}

\subsection{Fractional Token Contract Interface}

The Fractional Token Contract Interface represents the shared ownership in an NFT and extends the ERC20 interface. It defines standard operations such as minting new tokens upon NFT deposit, transferring tokens between owners, and burning tokens to adjust ownership shares or upon withdrawal of the NFT from the vault. Upon a successful NFT deposit, this function is invoked to mint the corresponding amount of fractional tokens, which are then allocated to the `to` address.

\textbf{a. Minting Fractional Tokens:}

Minting creates new fractional tokens, distributed to an NFT's depositor, representing their ownership share.

\begin{itemize}
    \item \textbf{to:} The address that will receive the minted tokens, typically the depositor's address. This parameter ensures the correct allocation of ownership shares in the form of fractional tokens.
    \item \textbf{amount:} Specifies the number of fractional tokens to mint, which correlates to the value of the deposited NFT.
\end{itemize}

\begin{lstlisting}[language=Solidity]
function mint(address to, uint256 amount) external;
\end{lstlisting}

\textbf{b. Transferring Fractional Tokens:}

The ability to transfer fractional tokens between addresses is fundamental for trading and redistributing ownership shares. 
This function facilitates the secure and flexible redistribution of fractional ownership, allowing token holders to engage in transactions that reflect changes in ownership interests.

\begin{itemize}
    \item \textbf{from:} The current holder's address of the fractional tokens. Ensuring that tokens are correctly debited from the rightful owner is crucial.
    \item \textbf{to:} The recipient's address of the tokens, enabling the transfer of ownership shares.
    \item \textbf{amount:} The number of tokens to be transferred, dictating the size of the ownership share being moved.
\end{itemize}

\begin{lstlisting}[language=Solidity]
function transfer(address from, address to, uint256 amount) external;
\end{lstlisting}

\textbf{c. Burning Fractional Tokens:}

Burning fractional tokens is a mechanism to reduce the total supply, typically used when an NFT is withdrawn. The burning process ensures the fractional token supply remains accurate and reflects the NFT's current fractional ownership distribution.

\begin{itemize}
    \item \textbf{from:} Indicates the address from which tokens will be burned, aligning the token supply with the current ownership structure.
    \item \textbf{amount:} The number of tokens to be burned directly impacts the representation of ownership shares within the ecosystem.
\end{itemize}

\begin{lstlisting}[language=Solidity]
function burn(address from, uint256 amount) external;
\end{lstlisting}

\subsection{Auction Contract Interface}

The Auction Contract facilitates competitive bidding processes for fractional tokens or entire NFTs. This contract manages auctions, tracks bids, determines winners, and ensures the transfer of ownership following auction conclusions.

\textbf{a. Initiating an Auction:}

To start an auction, the NFT owner or a fractional token holder specifies the asset to be auctioned, the starting price, and the auction duration.

\begin{lstlisting}[language=Solidity]
function startAuction(address assetAddress, uint256 tokenId, uint256 
startingPrice, uint256 duration) external;
\end{lstlisting}

- \textbf{assetAddress:} The contract address of the asset being auctioned, which could be the Vault Contract for an NFT.

- \textbf{assetId:} For an NFT, this is the tokenId; 
- \textbf{startingPrice:} The minimum bid required to participate in the auction.
- \textbf{duration:} The time frame for placing bids.

\textbf{b. Placing Bids:}

Participants can place bids on an active auction by sending the bid amount in the transaction, which must be higher than the current highest bid.

\begin{lstlisting}[language=Solidity]
function placeBid(uint256 assetId) external payable;
\end{lstlisting}

- \textbf{assetId:} The assetId of the item in the auction
- The sent value (\textit{msg.value}) must exceed the current highest bid for the bid to be considered valid.

\textbf{c. Concluding the Auction:}

After the auction duration has elapsed, the contract concludes the auction, transferring the auctioned asset to the highest bidder and redistributing the bid amounts accordingly. This function ensures that the highest bidder receives the asset, returning the bids to unsuccessful bidders and transferring the winning bid amount to the asset's seller.

\begin{lstlisting}[language=Solidity]
function endAuction(uint256 auctionId) external;
\end{lstlisting}

\textbf{d. Canceling an Auction:}

The Auction Contract also provides functionality to cancel an ongoing auction. This action can be initiated by the contract's governance mechanism under specific circumstances, such as detecting fraudulent activity or at the asset owner's request for legitimate reasons. Upon cancellation, the auction state is updated to reflect that it is no longer active, and any bids placed during the auction are refunded to the respective bidders.

\begin{lstlisting}[language=Solidity]
function cancelAuction(uint256 assetId);
\end{lstlisting}

\textbf{assetId:} The ID of the NFT whose auction is to be canceled.
This function SHOULD only be called by the governance contract or an account with equivalent permissions, ensuring that the cancellation process is controlled and secure.

\textbf{e. Redeeming Fractional Value:}

After the sale of an NFT, fractional token holders are entitled to redeem their share of the sale proceeds. The redeem function calculates the fractional tokens' value based on the NFT's total sale proceeds and transfers the corresponding Ether amount to the token holder.
Fractional tokens SHOULD be burned upon redemption to prevent double-spending.
\begin{lstlisting}[language=Solidity]
function redeemFractionValue(uint256 assetId, uint256 fractionAmount) external;
\end{lstlisting}

\textbf{assetId:} The ID of the asset whose sale proceeds are being claimed.
\textbf{fractionAmount:} The amount of fractional tokens the holder wishes to redeem.

\subsection{Market Integration Contract Interface}

The Market Integration Contract Interface streamlines decentralized finance (DeFi) ecosystem interactions for a broad spectrum of ERC20 tokens. This interface enables users to effectively manage liquidity pools, execute trades, and interface with various DeFi platforms, thereby enhancing the utility and liquidity of ERC20 tokens.

\textbf{a. Liquidity Management:}

This functionality permits users to add to or withdraw from the liquidity of any ERC20 token pair.

\begin{lstlisting}[language=Solidity]
function addLiquidity(uint256 tokenId, uint256 tokenAmount, 
uint256 pairTokenId, uint256 pairTokenAmount) external;

function removeLiquidity(uint256 tokenId, uint256 tokenAmount, 
uint256 pairTokenId, uint256 pairTokenAmount) external;
\end{lstlisting}

\begin{itemize}
    \item \textbf{tokenId, pairTokenId:} Identifiers for the ERC20 tokens involved in the liquidity transaction.
    \item \textbf{tokenAmount, pairTokenAmount:} Quantities of the respective ERC20 tokens to be added to or removed from the liquidity pool.
\end{itemize}

\textbf{b. Trade Execution:}

Enables swapping one ERC20 token for another within the pool to leverage market dynamics and diversify their asset holdings efficiently.

\begin{lstlisting}[language=Solidity]
function executeTrade(uint256 tokenIdIn, uint256 amountIn, uint256 tokenIdOut,
uint256 minAmountOut) external;
\end{lstlisting}

\begin{itemize}
    \item \textbf{tokenIdIn:} The identifier of the ERC20 token being offered in the exchange.
    \item \textbf{amountIn:} The volume of the input ERC20 token being traded.
    \item \textbf{tokenIdOut:} The identifier of the ERC20 token desired from the trade.
    \item \textbf{minAmountOut:} The minimum acceptable amount of the output ERC20 token ensures the trade is executed within acceptable slippage limits.
\end{itemize}

\textbf{c. Interfacing with External Platforms:}

The interface supports interactions with various DeFi platforms, enabling users to leverage their ERC20 tokens across multiple protocols for liquidity provision, trading, and other financial activities.

\subsection{Governance Contract Interface}

The Governance Contract for Fraction Holders is specifically designed to facilitate a democratic governance process. It allows fraction holders to propose, vote on, and implement changes. This system ensures that decision-making is aligned with the community's interests, leveraging ERC20 tokens for voting power.

\textbf{a. Proposal Creation and Management:}

Fraction holders can initiate governance proposals subject to a community vote. Proposals may include changes to protocol parameters, upgrades, or other significant decisions.

\begin{lstlisting}[language=Solidity]
function createProposal(string memory description, 
address target, bytes memory data) public;
\end{lstlisting}

- \textbf{Description:} A brief overview of the proposal's purpose and objectives.
- \textbf{Target:} The contract address that the proposal aims to interact with.
- \textbf{Data:} The encoded function calls data required for the proposal's execution.

\textbf{b. Voting Mechanism:}

The contract allows token holders to cast votes on active proposals. The voting power of each holder SHOULD be proportional to their token ownership, ensuring a fair and representative governance process.

\begin{lstlisting}[language=Solidity]
function vote(uint256 proposalId, bool support) public;
\end{lstlisting}

- \textbf{ProposalId:} The unique identifier of the proposal being voted on.
- \textbf{Support:} A boolean value indicating whether the vote is in favor of (true) or against (false) the proposal.

\textbf{c. Proposal Execution:}

After the conclusion of the voting period, proposals that meet the required quorum and approval thresholds are executed, enacting the proposed changes within the system. Execution is contingent upon meeting predefined conditions, such as quorum and majority support.

\begin{lstlisting}[language=Solidity]
function executeProposal(uint256 proposalId) public;
\end{lstlisting}

\subsection{Fee Management Contract Interface}

The Fee Management Contract Interface is designed to handle the financial aspects of the NFT fractionalization platform, explicitly focusing on the collection and distribution of fees. This contract ensures that fees generated from various operations within the ecosystem are managed transparently and efficiently.

\textbf{a. Collecting Fees:}

Fees are collected from various activities, such as NFT transactions, fractional token trades, and other services the platform provides. This function is responsible for accumulating user fees and securely storing them within the contract for future distribution or reinvestment into the platform.
 
\begin{lstlisting}[language=Solidity]
function collectFees() external payable;
\end{lstlisting}

\textbf{b. Distributing Fees:}

The collected fees are distributed according to predefined rules, including rewarding token holders, covering operational costs, or funding community proposals. This function outlines the mechanism for allocating the collected fees, ensuring stakeholders are compensated for their participation and investment in the platform.

\begin{lstlisting}[language=Solidity]
function distributeFees() external;
\end{lstlisting}

\subsection{Security and Functional Properties}
\label{sec:properties}

\textbf{a. Essential Properties:}

\label{subsec:essential_properties}
To ensure the platform's compatibility, security, and comprehensive functionality, the fractionalization abstraction adheres to the following properties, inspired by the principles underlying the ERC-721 \cite{erc721} and ERC-20 \cite{erc20} standards:

\begin{enumerate}

    \item \textbf{Asset Traceability}: Enable access to the associated NFT's address and tokenID for third parties, ensuring transparency and ease of integration with external platforms and services.
    
    \item \textbf{Proportional Gains Distribution}: Ensure proportional distribution of any revenue generated from the NFT based on the ownership share in the fractional tokens, promoting fairness and transparency in profit sharing.
    
    \item \textbf{Prevent Double Withdrawal}: Implement safeguards to prevent users from claiming their share of the proceeds more than once, protecting against double withdrawal vulnerabilities.
    
    \item \textbf{Governance Participation}: Allow fractional token holders to participate in governance decisions regarding the NFT, including its sale or use, reinforcing the decentralized nature of asset management.

    \item \textbf{Fraud and Tamper Resistance}: Incorporate security measures to protect against unauthorized access, fraud, and tampering, maintaining the integrity and trustworthiness of the fractionalization process.
\end{enumerate}

\textbf{b. Verification of Properties:}

Developers aiming to respect the outlined properties in Section \ref{subsec:essential_properties} should diligently implement the following verification, while users are encouraged to confirm these aspects before engaging with the solution:

\begin{enumerate}

    \item \textbf{Asset Traceability}: Stores the NFT address and ID within the smart contract, providing getter functions for external access, thus ensuring transparency regarding the underlying asset.
    
    \item \textbf{Proportional Gains Distribution}: Leverages the ERC-20 \emph{totalSupply} and \emph{balanceOf} functions to proportionally distribute gains among fractional token holders based on their ownership stake.
    
    \item \textbf{Prevent Double Withdrawal}: Ensure the smart contract executes the \emph{burn} function before sending gains to fractional holders during the redemption process. Burning tokens permanently removes them from circulation, eliminating the risk of withdrawing gains more than once.
    
    \item \textbf{Governance Participation}: Implements a voting system where fractional holders can vote on NFT management proposals, ensuring democratic decision-making processes. And create a modifier for setter functions that require governance.

    \item \textbf{Fraud and Tamper Resistance}: Incorporates security practices such as access controls, audit trails, and smart contract audits to safeguard against unauthorized actions and ensure the contract's integrity.
\end{enumerate}

These properties and their verification methods establish a secure and functional framework for NFT fractionalization, aiming to maximize compatibility with existing standards and protocols while providing a comprehensive set of asset management and trading features.

\section{Development of a Concrete Implementation}
\label{concrete}
In our proposed concrete implementation of the NFT fractionalization standard, we developed a set of smart contracts written in Solidity that facilitate the creation, management, and trading of fractional NFT shares within the Ethereum ecosystem. The complete code and tests are available on GitHub\footnote{https://github.com/yorku-ease/NFT-F}.

We combined the functionalities of \emph{IVault} and \emph{IAuction} into a single vault contract to reduce gas costs and simplify operations. Alongside this, we introduced a governance contract for decentralized governance and a simplified, automated market contract to support fractional shares trading. We integrated fee management directly into each contract, streamlining the architecture and allowing for customized fee strategies.

\subsection{Vault Smart Contract}

In the proposed implementation of the NFT fractionalization vault, we have integrated functionalities and security measures to support the fractional ownership of NFTs, decentralized auctions, and dynamic governance. 
The \emph{Vault} contract's key components and their functionalities are outlined as follows:

\begin{figure*}[ht]
    \centering
    \begin{subfigure}[b]{0.45\textwidth}
        \centering
        \includegraphics[width=\textwidth]{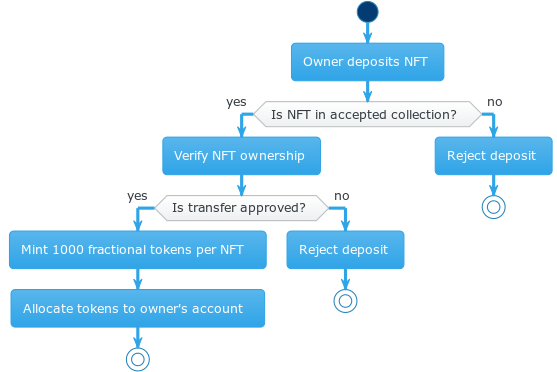}
        \caption{NFT management}
        \label{fig:act-management}
    \end{subfigure}
    \hfill
    \begin{subfigure}[b]{0.45\textwidth}
        \centering
        \includegraphics[width=\textwidth]{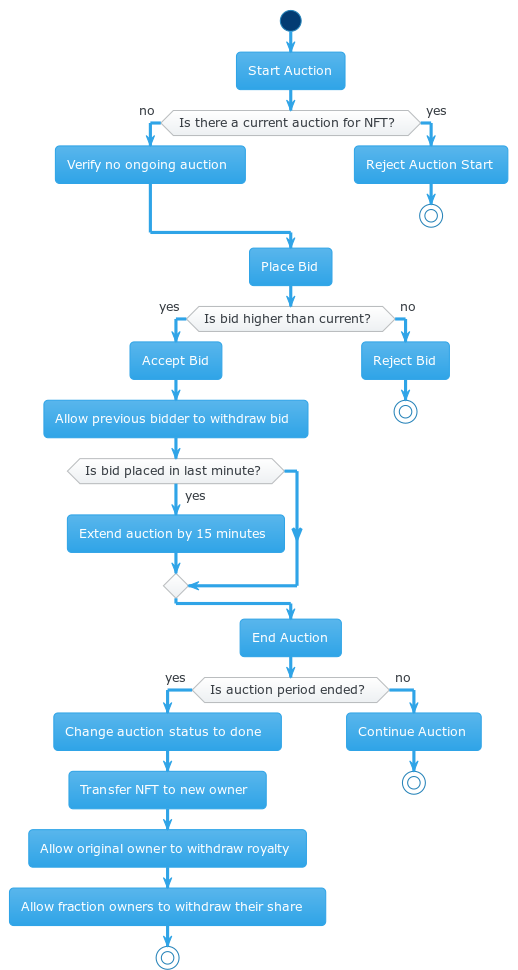}
       
        \caption{Auction management}
        \label{fig:act-auction}
    \end{subfigure}
    \caption{Vault Activity diagrams}
    \label{fig:act-vault }
\end{figure*}

\begin{itemize}
    \item \textbf{NFT Management:} The NFT management process is presented in Figure \ref{fig:act-management}. The vault facilitates the fractionalization process by enabling users to deposit and withdraw NFTs. It accepts multiple ERC721 NFTs, each belonging to the pre-specified NFT collection determined at the contract's initiation. Upon deposit, fractional tokens are minted and allocated to the depositor, correlating each NFT to 1000 fractional tokens. In this example, the number of tokens is fixed for simplicity. Future implementations could explore optimal fractionalization rates depending on the type of asset and its market demand. This mechanism simplifies the management and trading of fractions, enhancing liquidity. The exact amount of fractional tokens corresponding to an NFT (1000) must be burned for withdrawals, establishing a clear link between NFT ownership and fractional token holdings.

    \item \textbf{Auction Management:} The auction management process is presented in Figure \ref{fig:act-auction}. We incorporate a voting auction mechanism for dynamic price discovery, where participants vote on the price, and the highest bid secures the NFT \cite{Wejdene}. Our implementation includes an \emph{Auction} struct to manage the state of NFT auctions within the vault. It lets us track the active status, end time, highest bid, and other relevant auction details. The default duration for auctions is seven days and can be updated with governance voting. The auction is initiated via the \emph{startAuction} function with specific parameters such as the NFT's asset address, token ID, starting price, and duration. The auction integrity is upheld by ensuring that bids exceed all previous offers and are placed before the auction concludes. Notably, if a bid is made within a predefined period before the auction's end, the duration automatically extends by 15 minutes to ensure fairness and prevent bid sniping.

    \item \textbf{Governance Functions:} The vault is controlled by a governance contract, allowing parameter updates. The governance features are enhanced by enabling the governance contract address to be set post-deployment through the \emph{setGovernanceContract} function, callable only once to prevent misuse. This governance contract can adjust auction durations, modify royalty percentages, and cancel auctions, adapting to market dynamics and community preferences. Using the \emph{onlyGovernance} modifier ensures that only authorized calls from the governance contract can execute certain functions.

    \item \textbf{Security Considerations:} Security is prioritized through various strategies, including the use of a reentrancy guard to protect against attacks, the implementation of the checks-effects-interactions pattern to prevent reentrancy issues, and access control modifiers to limit operations to authorized entities.
\end{itemize}

\subsection{FractionalToken Smart Contract}
In the proposed implementation, we introduce the  \emph{FractionalToken}  contract, an ERC20 token representing fractional ownership of Non-Fungible Tokens (NFTs). The  \emph{FractionalToken}  contract implements the \emph{IFractionalToken} interface, which extends the ERC20 token standard, adhering to the proposed NFT fractionalization framework. This ensures that it complies with both the ERC20 standard and the additional functionalities specific to fractional ownership.

 This contract is implemented to incorporate minting and burning functionalities, essential for managing the dynamic supply of fractional tokens following the lifecycle of NFTs within a vault. 

The \emph{FractionalToken} contract is associated with an NFT vault, indicated by the \emph{nftVault} address variable. The NFT vault address is authorized to mint or burn the fractional tokens. We have implemented a custom modifier, \emph{onlyVault}, to enforce that only the designated NFT vault can execute minting and burning operations. 
Upon deployment, the contract's constructor initializes the token with a specified name and symbol and designates the deploying address as the contract owner. 

The \emph{mint} function facilitates the issuance of new tokens to represent fractional ownership shares of NFTs deposited in the vault. Conversely, the \emph{burnFrom} function reduces fractional tokens in circulation, mirroring the withdrawal or reallocation of NFT assets.

\subsection{GovernanceContract Smart Contract}

In our proposed implementation, the \emph{GovernanceContract} facilitates governance for fraction holders, utilizing an ERC20 token representing fractional ownership as the foundation for voting. This approach ensures that decision-making power is proportionally distributed according to fractional ownership. To enhance the security and integrity of the governance process, we integrated the \emph{TimelockController} from OpenZeppelin, introducing a mandatory delay between the approval of proposals and their execution.

\begin{figure*}[ht]
\centering
\includegraphics[width=1\textwidth]
{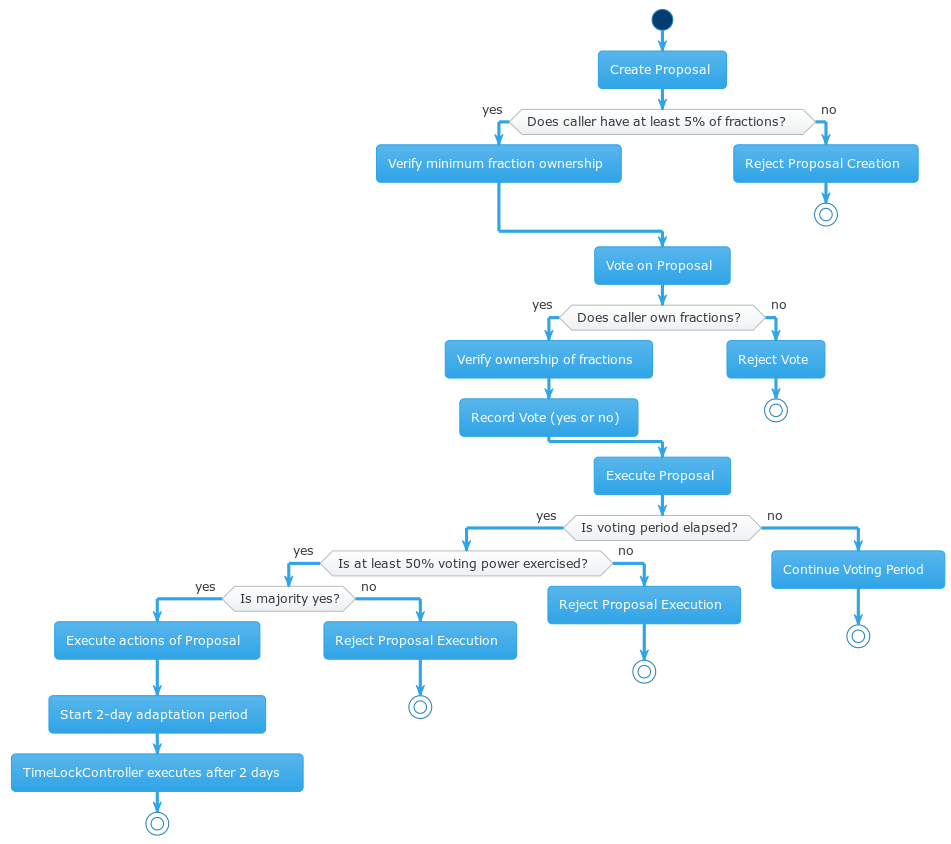}
\caption{Governance mechanism}
 \label{fig:act-gov}
\end{figure*}

Figure \ref{fig:act-gov} presents the activity diagram of the governance process. Proposals, the core of the governance mechanism, are structured to include necessary details such as a descriptive goal, the target contract address for execution, encoded function call data, and the defined voting period. Each proposal also tracks its execution status, votes for and against, total votes cast, and the total token supply at creation to determine a quorum.

In our design, initiating a proposal is conditioned upon the proposer holding a minimum threshold of the total token supply. This ensures that only stakeholders with a significant interest can propose governance actions. A quorum threshold, set at 50\% of the total token supply, is required for a proposal to be considered valid.

The voting process we implemented allows token holders to cast their votes on active proposals within a designated voting period. The process is secured against reentrancy attacks to maintain its integrity. This design ensures that governance actions reflect the collective will of the fraction holders.

For a proposal to proceed to execution, it must conclude the voting period, achieve the necessary quorum, and receive a majority of votes in favor. Approved proposals are then scheduled for execution through the \emph{TimelockController} after a specified delay, providing a window for review and potential intervention, enhancing the governance model's transparency and security.

\subsection{MarketIntegration Smart Contrac}

The \emph{MarketIntegration} contract aims to provide a simple solution for managing liquidity and enabling trades between two distinct ERC20 tokens. These tokens may represent varied assets, such as fractions of ownership in distinct NFT vaults or a mix of a fractional token and a standard ERC20 cryptocurrency like Ether. This versatility underscores the contract's utility in a broad spectrum of decentralized finance (DeFi) applications; the contract illustrates the underlying mechanics of liquidity provision and token trading in a simplified manner.

\begin{figure*}[ht]
\centering
\includegraphics[width=0.8\textwidth]
{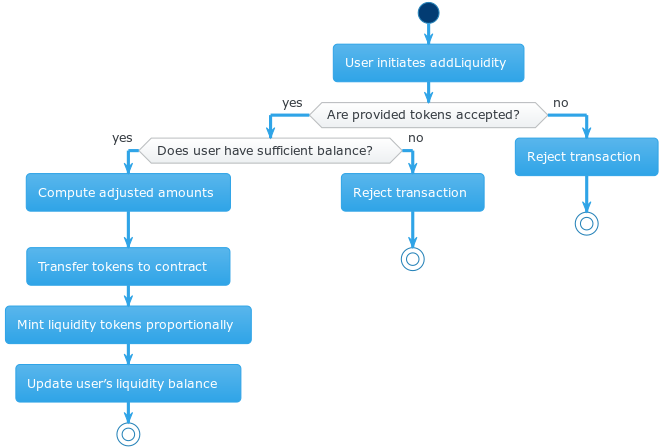}
\caption{Add Liquidity}
 \label{fig:act-add}
\end{figure*}

\begin{figure*}[ht]
\centering
\includegraphics[width=0.6\textwidth]
{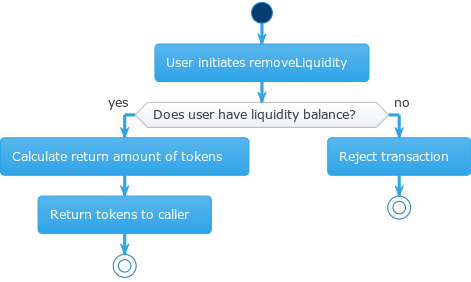}
\caption{Remove Liquidity}
 \label{fig:act-remove}
\end{figure*}

\begin{figure*}[ht]
\centering
\includegraphics[width=1\textwidth]
{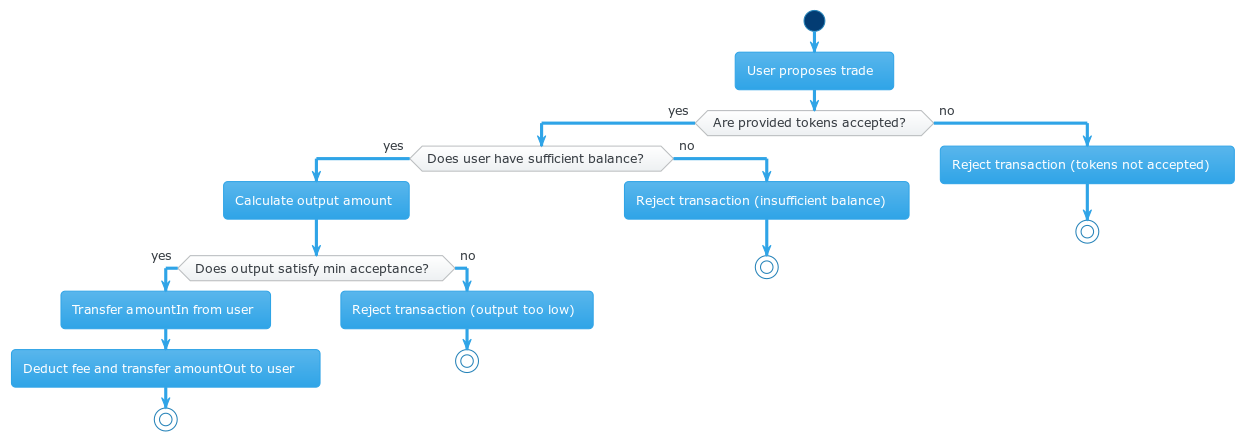}
\caption{Trade Liquidity}
 \label{fig:act-trade}
\end{figure*}

Upon initialization, the contract requires the addresses of two ERC20 tokens. These addresses enable the contract to interact with assets representing ownership in various assets or a pairing of investment and utility tokens. 
\ref{fig:act-add} presents the activity diagram for adding liquidity, whereas \ref{fig:act-remove} presents the activity diagram for removing liquidity. The addition and removal of liquidity are straightforward. Users can deposit or withdraw their tokens into or from the liquidity pool, figure 

Trading between the two tokens is facilitated by a mechanism that considers the current liquidity reserves and employs slippage protection. The trading flow is presented is figure \ref{fig:act-trade}. lippage refers to the difference between the expected price of a trade and the price at which the trade is executed. It occurs due to price movement between when a transaction is initiated and when it is completed, often exacerbated in volatile market conditions. Our contract addresses slippage by allowing traders to specify a minimum acceptable amount of output tokens, minAmountOut, ensuring traders receive a value close to their expectation or the transaction fails, thus protecting them from unfavorable price movements. 

The contract uses a straightforward formula to calculate the amount out for a trade, factoring in the input amount, liquidity reserves, and a predefined trading fee. Specifically, we assume a nominal fee of 0.3\%, deducted to incentivize liquidity provision and mitigate potential arbitrage. 

The output amount ($amountOut$) for a given input amount ($amountIn$) in a trade is calculated using the formula:
\[
amountOut = \frac{amountInWithFee \times outputReserve}{(inputReserve \times 10000) + amountInWithFee}
\]
Where:
\begin{itemize}
    \item $amountInWithFee = amountIn \times fee$
    \item $fee = 9975$, representing a 0.3\% trading fee ($10000 - fee$ gives the fee percentage).
    \item $inputReserve$ and $outputReserve$ are the liquidity reserves for the input and output tokens, respectively.
\end{itemize}


\subsection{Deployment of the concrete Implementation}
Our smart contracts are deployed within the Truffle Development Environment \cite{truffle}, activated through the \emph{Truffle develop} command. This command initiates a local Ethereum blockchain simulation for development and testing, featuring instant transaction mining and a set of pre-funded accounts.

For contract deployments and interactions, we used account 0, the first in the array of automatically generated accounts, as the transaction sender.

The deployment begins with the \emph{FractionalToken} contract, labeled as "FTK," which plays a dual role within our ecosystem. As a fractional ownership token, FTK represents shares in the NFTs held within the vault, and as the governance token, grants holders the right to vote on proposals affecting the ecosystem's operations and the management of the vaulted NFTs. 

Following the FTK token deployment, we deploy the \emph{MyToken} contract. This contract is the collection accepted by the vault. Each NFT minted from \emph{MyToken} can be deposited into the vault for fractionalization, where FTK tokens represent it.

The \emph{Vault} contract is deployed next, configured to recognize \emph{MyToken} as its accepted NFT collection and utilizing FTK for the fractionalization process. 

To enhance the security and integrity of the governance process, the \emph{TimelockController} contract is then deployed. This contract introduces a mandatory delay between the approval of proposals and their execution.

The \emph{GovernanceContract} is deployed and utilizes FTK as the voting token. This contract oversees the proposal and voting system, allowing FTK holders to influence the ecosystem's direction and decisions. Upon deployment, the GovernanceContract's address is registered with the \emph{Vault}, integrating the governance functionalities with the vault's operations.

Finally, the \emph{MarketIntegration} contract is deployed to enable liquidity and trading between FTK and another ERC-20 token, identified here as "TB." 


\subsection{Security Analyse}
For the security analysis of our implementation, we employed several tools, including Slither \cite{slither} for static analysis, Mythril \cite{mythril} for symbolic execution, and Echidna \cite{echidna} for property-based testing to identify potential vulnerabilities and ensure the robustness of our smart contracts.

\subsubsection{Slither}

After compiling our smart contracts in the truffle environment, we utilized Slither to conduct a comprehensive security assessment. By running \emph{slither .} at the project's root directory, we enabled Slither to examine all Solidity files for potential vulnerabilities. 
The outcome of Slither's examination revealed that four high-severity issues, two medium-severity issues, and several minor and informational errors were reported. Below, we detail each reported error alongside the solutions we opted for.

\textbf{High Severity Issues:}

\begin{enumerate}
   \item Slither highlighted security risks in the \emph{Vault.endAuction} and \emph{Vault.redeemFractionValue} functions related to direct Ether transfers to arbitrary addresses. Such transfers can lead to reentrancy attacks or unintended code execution if the recipient is a malicious contract.

We adopted the "Pull Over Push" strategy to mitigate these risks, significantly reducing our contract's exposure to such threats. This approach involves:

\begin{itemize}
     \item Creating a Pending Withdrawals Ledger: We introduced a mapping to track funds due to users, thus decoupling fund accumulation from withdrawal.
     \item Implementing a Secure Withdraw Function: We developed a withdraw function that lets users safely retrieve their funds, ensuring atomicity to prevent reentrancy attacks.
      \item Ensuring Graceful Error Handling: The withdraw function is designed to revert transactionally if Ether transfers fail, safeguarding the contract's integrity.
   \end{itemize}
   
   \item Slither's analysis identified potential vulnerabilities in the \emph{TimelockController.\_execute} and 
   
   \emph{Vault.redeemFractionValue} functions due to their use of call operations for sending Ether (ETH) to arbitrary addresses.

However, we've determined this concern to be a false positive in our context. The \emph{Vault.redeemFractionValue} function's design, which allows users to initiate withdrawals, already mitigates the risk of reentrancy by following secure patterns. Despite Slither's warnings, our analysis concludes that the safeguards, such as the non-reentrant modifier and user-initiated withdrawal mechanisms, effectively minimize the risks associated with these operations.

\item Slither's examination of our MarketIntegration contract revealed an oversight in handling ERC20 transfer and \emph{transferFrom} methods, where we didn't check their return values. This oversight could lead to inaccuracies in token balance accounting, posing a risk of unintended contract behaviors and vulnerabilities since these methods are expected to return a boolean indicating success or failure.

To mitigate this, we've updated our contract to rigorously check the return values from all ERC20 token operations, employing required statements to ensure their success.

\item Slither highlighted a concern with our FractionalToken contract, specifically regarding the uninitialized state variable \emph{isNftVaultSet}. This variable, intended to indicate if the NFT vault address has been set, was reported as never being initialized, yet it's referenced in the \emph{updateNFTVault} function. 
To resolve this, we've ensured that \emph{isNftVaultSet} is properly initialized within our contract during contract initialization.

\end{enumerate}

\textbf{Medium Severity Issues:}

\begin{enumerate}
\item Slither flagged strict equality checks in OpenZeppelin's \emph{TimelockController.getOperationState} function as potential issues, specifically for comparisons with 0 and \_DONE\_TIMESTAMP. These checks are vital for determining the operation's state but can pose risks if not carefully managed.

This concern, however, is identified as a false positive. OpenZeppelin's design choice to use strict equality checks is intentional and necessary for the function's accurate operation management. While Slither's alert highlights the importance of cautious implementation, the checks are appropriate and secure in this context. 

\item Slither identified potential reentrancy vulnerabilities in the \emph{Vault.redeemFractionValue}, \emph{MarketIntegration.removeLiquidity}, and \emph{Vault.withdrawNFT} functions of our smart contracts. These issues stem from making external calls before updating state variables, which could allow for reentrancy attacks.

We've implemented the Checks-Effects-Interactions pattern across our contracts to address these concerns.

\end{enumerate}

\textbf{Low Severity Issues and Information:}

\begin{enumerate}
    \item \textbf{Variable Shadowing in Constructor Parameters}: In the \emph{FractionalToken}, constructor parameters \emph{name} and \emph{symbol} were shadowing inherited \emph{ERC20} and \emph{IERC20Metadata} functions. To eliminate shadowing and clarify the code, the parameters were renamed to \emph{\_name} and \emph{\_symbol}.

    \item \textbf{Absence of Events for State Changes}: The \emph{FractionalToken.updateNFTVault} function made state changes without event emission, obscuring transparency. A new event, \emph{NFTVaultUpdated}, was introduced and emitted whenever \emph{nftVault} is updated, enhancing contract transparency.

    \item \textbf{Missing Zero Address Validation}: Functions like \emph{Vault.setGovernanceContract} lacked validation for zero address inputs, posing a risk of logical errors. This issue was mitigated by adding \emph{require} statements to ensure non-zero addresses, preventing inadvertent assignment of zero addresses.

    \item \textbf{External Calls Inside Loops}: In functions such as \emph{TimelockController.\_execute} and \emph{Vault.depositNFTs}, making external calls within loops raised gas and security concerns. The functions were restructured to minimize external calls within loops, ensuring gas efficiency and security.

    \item \textbf{Reentrancy Vulnerabilities}: Functions including \emph{MarketIntegration.addLiquidity} were identified as vulnerable to reentrancy attacks due to state updates after external calls. Adopting the \emph{nonReentrant} modifier and restructuring functions to ensure state updates precede external interactions mitigated these reentrancy risks.

    \item \textbf{Reliance on \emph{block.timestamp}}: Several functions relied on \emph{block.timestamp} for critical comparisons, which miners can slightly manipulate. The contract logic was reviewed to ensure it accounts for minor variations and potential manipulations, mitigating this issue.

    \item \textbf{Usage of Low-Level Calls}: The use of low-level calls in functions like \emph{TimelockController.\_execute} posed security risks. However, within OpenZeppelin's \emph{TimelockController}, this is a necessary and deliberate choice to provide flexibility in executing arbitrary transactions. These instances are carefully implemented to manage potential reentrancy risks and ensure proper error handling.

    \item \textbf{Similar Variable Names}: Variables with similar names across \emph{MarketIntegration} and interfaces could cause confusion. By renaming variables for clarity and adopting distinct naming, this issue was mitigated to prevent confusion and improve code readability.

    \item \textbf{Non-Constant State Variables}: Some state variables could have been declared as \emph{constant} or \emph{immutable} for gas optimization. This was addressed by reviewing and marking applicable variables as \emph{immutable} or \emph{constant}, optimizing gas usage and contract efficiency.

    \item \textbf{Solidity Version Alignment}: Concerns were raised about the use of the latest Solidity version \emph{\^{}0.8.20} being potentially too recent. However, adopting the latest Solidity version by OpenZeppelin contracts aims to leverage improvements and security enhancements introduced in newer compiler versions.

    \item \textbf{Inline Assembly Usage}: The presence of inline assembly in \emph{Address.sol} was flagged for potential security risks. Inline assembly within OpenZeppelin's utilities, such as \emph{Address.sol}, is judiciously used to perform operations not feasible with Solidity's high-level language alone. These usages are critical for certain functionalities, such as reliably detecting contract addresses, and are implemented with utmost care to maintain security.
\end{enumerate}

\subsubsection{Mythril}
After conducting a comprehensive security analysis using Mythril on the four smart contracts, the results indicated no vulnerabilities or issues were detected. This outcome underscores the effectiveness of the coding practices adopted and the fixes implemented after the Slither analysis.

\subsubsection{Echidna}

We employed the Echidna testing framework to conduct an in-depth examination of the properties of our proposed concrete implementation. Below is a summary of the tests carried out; all were successfully passed.

\begin{itemize}
 \item \textbf{FractionalToken Contract Testing:}
 Echidna tests for the \emph{FractionalToken} contract focus on essential operations such as minting, burning, and supply management; 

\textit{Minting Authorization Test}: Assesses access controls for the minting function, ensuring only authorized entities can mint tokens, thereby preventing unauthorized supply inflation.
    
 \textit{Burning Authorization Test}: Evaluates the burning function's access restrictions, confirming that only specified roles can execute burns to maintain supply integrity.
    
  \textit{Total Supply Invariance Test} 
    
    (\emph{echidna\_test\_total\_supply\_constant}): Tests the contract’s capability to maintain a constant total supply in the absence of minting or burning activities.

\item \textbf{Vault Contract Testing:}
Testing of the \emph{Vault} contract within the Echidna suite evaluates governance controls, auction mechanics, and asset management:


  \textit{Governance Contract Singleton Test} : Ensures the governance contract address can be set only once, cementing governance stability.
    
\textit{Positive Auction Duration Test}: Verifies that auction durations are set above zero, ensuring auctions occur over a meaningful timeframe.
    
\textit{Royalty Percentage Range Test}: Confirms that the royalty percentage falls within a viable range (0-100\%), maintaining fairness and viability in royalty mechanisms.
    
  \textit{Withdrawal Balance Check}: Validates that withdrawals cannot proceed without sufficient balance, safeguarding against unauthorized fund extractions.
    
       \item \textbf{Correct Original Owner on Deposit Test} : Ensures the original owner is correctly recorded upon NFT deposit, crucial for ownership integrity and royalty distributions.
 \item \textbf{Governance Mechanism Integrity:}
 Focuses on the \emph{GovernanceContract}'s functionality, specifically proposal creation, voting mechanisms, and execution of governance actions:
 

\begin{itemize}
    \item \textbf{Voting Increases Votes Test} 
    
    (\emph{echidna\_test\_voting\_increases\_votes}): Tests the fundamental aspect of voting, ensuring that casting a vote correctly increases the proposal's vote count, validating the vote recording mechanism.
    
    \item \textbf{Quorum Not Met Test} 
    
    (\emph{echidna\_test\_quorum\_not\_met}): Ensures proposals without sufficient voter turnout are not incorrectly approved, maintaining the requirement for a minimum level of community engagement for governance decisions.
    
    \item \textbf{Create Proposal Functionality Test} 
   
    (\emph{echidna\_test\_create\_proposal}): Verifies the proposal creation process, ensuring that the system can dynamically add new proposals under appropriate conditions, keeping the governance system responsive and inclusive.
\end{itemize}
\item \textbf{MarketIntegration Contract Examination:}
The \emph{MarketIntegration} contract's Echidna tests validate liquidity management, trade functionality, and supply conservation: 


\begin{itemize}
    \item \textbf{Liquidity Maintenance Test} 
    
    (\emph{echidn\_test\_liquidity\_maintenance}): Evaluates the contract's ability to correctly manage liquidity levels, ensuring that liquidity pools are not unintentionally depleted. This test simulates various liquidity scenarios to affirm the contract's resilience and capacity to sustain market operations.
    
    \item \textbf{Trade Execution Verification Test} 
    
    (\emph{echidna\_test\_trade\_execution}): Assesses the contract's trade execution logic, ensuring that trades are processed only if they meet specific conditions, such as slippage tolerance and minimum liquidity requirements. This safeguards the market from potential manipulations and ensures the integrity of trade transactions.

\item \textbf{Supply Management Accuracy Test} 

(\emph{echidna\_test\_supply\_management}): Confirms the contract's effectiveness in managing token supply accurately during liquidity events. It tests the contract's response to liquidity addition and removal, verifying that supply changes reflect actual token dynamics without unintended discrepancies.

\end{itemize}

\end{itemize}

\section{Related Work}
\label{Related}

The academic exploration of fractional NFTs remains limited, with most insights coming from industry reports rather than scholarly research. 

Haouari et al. \cite{Wejdene} explore the concept of NFT fractionalization and abstract smart contracts for fractionalization smart contracts. Their study examines the smart contracts used by platforms such as Fractional.art and Unicly, to establish best practices and recommendations for developing secure and efficient fractionalization protocols. While the authors provide a foundational abstraction for fractionalization contracts, their proposed platform lacks critical mechanisms, such as integrated governance structures. Additionally, the study does not include a comprehensive security analysis or provide a concrete implementation of the proposed standard, leaving important aspects of practical application and security validation unaddressed.

Choi et al. \cite{Wonseok} explore fractional non-fungible tokens (NFTs), emphasizing their role in democratizing access to high-value digital assets through blockchain technology and tokenization. Their paper provides a foundational overview of tokenization, focusing on the potential and challenges of fractional NFTs. Although the study delves into the mechanisms of creation, management, and existing platforms for fractional NFTs, as well as gas fees, it notably does not examine the smart contracts of each solution in detail, particularly their security mechanisms.

Choi et al. \cite{10174920} present a detailed analysis of gas costs associated with the implementation of fractional NFTs on the Ethereum blockchain. The study explores ERC-20, ERC-721, and ERC-1155, which are essential for minting fractional NFTs. By examining the gas consumption of each standard during the deployment, minting, approval, and transfer processes, the authors identify that while ERC-1155 shows efficiency in specific stages, its overall gas cost is higher compared to ERC-721 when multiple types of tokens are minted and transferred simultaneously. However, the study focuses on the gas costs of minting a single NFT in each smart contract, which may not reflect the complexities and costs of real-world applications where multiple NFTs are minted and traded simultaneously.  Also the study does not consider how differences in implementation across various platforms might affect the analysis, potentially leading to varying gas costs and performance metrics.

Rudytsia and Bogdanova \cite{Rudytsia_2022} introduce a UML model for NFT fractionalization focusing on technical framework development for blockchain-based property rights distribution, adhering to the ERC-1155 standard. The paper provides a solution for creating, selling, and managing fractional NFTs. However, it does not address the security implications of their proposed model, lacks a buyout option, and specifies that only non-fractionalized NFTs can be purchased, potentially limiting the practicality and adaptability of their approach in diverse application scenarios.

 Fang from \emph{CoinGecko} \cite{coingecho} explores the concept of fractionalized NFTs, focusing on platforms like Unic.ly and NIFTEX. They discuss how fractionalization can enhance accessibility and liquidity in the NFT market. However, the analysis would benefit from updating certain information, adding technical details, and including a security assessment to provide a better understanding of the risks and technical intricacies involved in fractionalizing NFTs. 

Vijayakumaran \cite{Vijayakumaran2021} analyzes the legal landscape surrounding fractionalized NFTs (F-NFTs), mainly focusing on security law implications. While their exploration offers valuable insights into regulatory considerations, it notably lacks an in-depth discussion on the technical processes and mechanisms underlying F-NFT fractionalization.

\section{Discussion and Future Work}
\label{future}
 While current platforms offer various methods for fractionalizing NFTs, the lack of a unified standard introduces significant security, interoperability, and scalability challenges. This paper proposes 
 standardization framework that aims to address these issues by offering a structured approach that ensures consistency and security across different platforms.

One of the primary benefits of standardization is the potential to enhance interoperability. By establishing a standard set of guidelines and protocols, different platforms can interact more seamlessly, allowing for the transfer and trading of fractionalized NFTs across various ecosystems. This would improve NFTs' liquidity and foster innovation by enabling developers to build on a shared foundation rather than reinventing the wheel for each new project.

Security is another critical area where standardization can have a profound impact. The variability in implementing fractionalization leads to inconsistent security practices, leaving assets vulnerable to attacks. A standardized approach would enforce best practices in smart contract design, access control, and transaction validation, reducing the likelihood of security breaches.

To realize the full potential of the proposed standard, further work is needed to refine and advocate for its adoption within the broader Ethereum community. Engaging with developers, platform operators, and other stakeholders will be crucial in ensuring that the standard meets the needs of the industry. Additionally, research into optimizing the efficiency of the proposed standard, particularly in terms of gas costs and performance, will be essential to making fractionalization accessible and scalable.

\section{Conclusion}
\label{Conclusion}

In this paper, we explored the concept of NFT fractionalization from both theoretical and practical viewpoints, proposing an abstract framework for smart contracts in this area. By clarifying the roles, interactions, and technical requirements involved in fractionalization, we aimed to foster a more inclusive digital asset ecosystem.
A crucial part of our research involved conducting a detailed security analysis and audit of smart contracts from platforms such as Tessera, NFTX, and Unicly. Tools like Slither and Mythril revealed no significant errors, attesting to the solidity of these platforms’ contracts. Our audit pinpointed common vulnerabilities, particularly in access control and validation, underscoring the need for focused improvements. Additionally, we illustrated the viability of our standardization proposal with a concrete implementation, emphasizing the importance of rigorous security testing to affirm the strength of such frameworks.

\bibliographystyle{ieeetr}

\bibliography{cas-refs}

\begin{thebibliography}{10}

\bibitem{wang2021nonfungible}
Q.~Wang, R.~Li, Q.~Wang, and S.~Chen, ``Non-fungible token (nft): Overview, evaluation, opportunities and challenges,'' 2021.

\bibitem{Wejdene}
H.~Wejdene and F.~Marios, ``Towards a standardization of fractionalization smart contracts for non-fungible tokens,'' in {\em Proceedings of the 33rd Annual International Conference on Computer Science and Software Engineering}, CASCON '23, (USA), p.~132–141, IBM Corp., 2023.

\bibitem{eip}
M.~Becze and H.~Jameson, ``Ethereum improvement proposals.'' Ethereum Improvement Proposals, 2015.

\bibitem{erc20}
F.~Vogelsteller and V.~Buterin, ``Erc-20: Token standard.'' Ethereum Improvement Proposals, 2017.

\bibitem{erc721}
W.~Entriken, D.~Shirley, J.~Evans, and N.~Sachs, ``Erc-721: Non-fungible token standard.'' Ethereum Improvement Proposals, 2017.

\bibitem{erc1155}
W.~Radomski, A.~Cooke, P.~Castonguay, J.~Therien, E.~Binet, and R.~Sandford, ``Erc-1155: Multi token standard.'' Ethereum Improvement Proposals, 2018.

\bibitem{erc165}
C.~Reitwießner, N.~Johnson, F.~Vogelsteller, J.~Baylina, K.~Feldmeier, and W.~Entriken, ``Erc-165: Standard interface detection.'' Ethereum Improvement Proposals, 2018.

\bibitem{Mohan2022}
V.~Mohan, ``Automated market makers and decentralized exchanges: a defi primer,'' {\em Financial Innovation}, vol.~8, no.~1, p.~20, 2022.

\bibitem{sushiswap}
uniswap, ``Uniswap protocol.'' \url{https://uniswap.org/}, 2023.
\newblock Accessed: 2023-11-23.

\bibitem{Tessera-medium}
B.~Galang, ``Getting started with tessera.'' \url{https://medium.com/tessera-nft/getting-started-with-tessera-c9709293ee88}, 2022.
\newblock Accessed: 2023-02-01.

\bibitem{Hyperstructures}
jacob, ``Hyperstructures.'' \url{https://jacob.energy/hyperstructures.html}, 2022.
\newblock Accessed: 2024-01-28.

\bibitem{tesseraGithub}
s.~S. andy8052, ``Fractional contracts.'' \url{https://github.com/code-423n4/2022-07-fractional}.
\newblock Accessed: 2023-10-23.

\bibitem{Tesseradown}
binance, ``Paradigm-backed nft ownership platform tessera is shutting down.'' \url{https://www.binance.com/en/square/post/518314}.
\newblock Accessed: 2024-06-06.

\bibitem{unicly}
unicly, ``unic.ly.'' \url{https://www.app.unic.ly/#/}, 2023.
\newblock Accessed: 2023-02-10.

\bibitem{mohan2022automated}
V.~Mohan, ``Automated market makers and decentralized exchanges: a defi primer,'' {\em Financial Innovation}, vol.~8, no.~1, p.~20, 2022.

\bibitem{nftxDoc}
NFTX, ``Introduction to nftx.'' \url{https://docs.nftx.io/}, 2022.
\newblock Accessed: 2023-05-01.

\bibitem{truffle}
ConsenSys, ``Truflle suite.'' \url{https://archive.trufflesuite.com//}.
\newblock Accessed: 2023-12-05.

\bibitem{slither}
J.~Feist, G.~Grieco, and A.~Groce, ``Slither: A static analysis framework for smart contracts,'' in {\em 2019 {IEEE}/{ACM} 2nd International Workshop on Emerging Trends in Software Engineering for Blockchain ({WETSEB})}, {IEEE}, may 2019.

\bibitem{mythril}
B.~Mueller, ``Smashing ethereum smart contracts for fun and real profit.'' \url{https://conference.hitb.org/hitbsecconf2018ams/materials/WHITEPAPERS/WHITEPAPER\%20-\%20Bernhard\%20Mueller\%20-\%20Smashing\%20Ethereum\%20Smart\%20Contracts\%20for\%20Fun\%20and\%20ACTUAL\%20Profit.pdf}, 2018.

\bibitem{echidna}
G.~Grieco, W.~Song, A.~Cygan, J.~Feist, and A.~Groce, ``Echidna: Effective, usable, and fast fuzzing for smart contracts,'' in {\em Proceedings of the 29th ACM SIGSOFT International Symposium on Software Testing and Analysis}, ISSTA 2020, (New York, NY, USA), p.~557–560, Association for Computing Machinery, 2020.

\bibitem{Wonseok}
W.~Choi, J.~Woo, and J.~Hong, ``Fractional non‐fungible tokens: Overview, evaluation, marketplaces, and challenges,'' {\em International Journal of Network Management}, 01 2024.

\bibitem{10174920}
W.~Choi, J.~Woo, and J.~W.-K. Hong, ``Gas cost analysis of fractional nft on the ethereum blockchain,'' in {\em 2023 IEEE International Conference on Blockchain and Cryptocurrency (ICBC)}, pp.~1--6, 2023.

\bibitem{Rudytsia_2022}
N.~B. Yehor~Rudytsia1, ``Uml model of the property right distribution module using nft fractionalization based on blockchain technology,'' {\em International Science Journal of Engineering and agriculture}, vol.~1, no.~3, p.~98–109, 2022.

\bibitem{coingecho}
L.~Fang, ``Fractionalized nft.'' CoinGecko, 2021.

\bibitem{Vijayakumaran2021}
A.~Vijayakumaran, ``Democratizing nfts: F-nfts, daos and securities law,'' {\em Richmond Journal of Law and Technology}, vol.~2021, November 2021.
\newblock Available at SSRN: https://ssrn.com/abstract=3964905 or http://dx.doi.org/10.2139/ssrn.3964905.

\end{thebibliography}
\end{document}